\documentclass[a4paper,twocolumn,prl,superscriptaddress,showpacs]{revtex4}
\usepackage{dcolumn}
\usepackage{amsmath}
\usepackage{graphicx}
\usepackage{latexsym}  
\usepackage{amsfonts}  
\usepackage{amssymb}  
\usepackage{color}
\DeclareGraphicsExtensions{.pdf,.png,.gif,.jpg}

\newcommand{\be}{\begin{equation}}
\newcommand{\ee}{\end{equation}}
\newcommand{\bea}{\begin{eqnarray}}
\newcommand{\eea}{\end{eqnarray}}

\begin{document}

\title{Dynamical Coulomb Blockade and the Derivative Discontinuity of Time-Dependent 
Density Functional Theory} 
 
\author{S. Kurth}  
\affiliation{Nano-Bio Spectroscopy Group, 
Dpto. de F\'{i}sica de Materiales, 
Universidad del Pa\'{i}s Vasco UPV/EHU, Centro F\'{i}sica de Materiales 
CSIC-UPV/EHU, Av. Tolosa 72, E-20018 San Sebasti\'{a}n, Spain} 
\affiliation{IKERBASQUE, Basque Foundation for Science, E-48011 Bilbao, Spain}
\affiliation{European Theoretical Spectroscopy Facility (ETSF)}
  
\author{G. Stefanucci}
\affiliation{Dipartimento di Fisica, Universit\`{a} di Roma Tor Vergata,
Via della Ricerca Scientifica 1, 00133 Rome, Italy}
\affiliation{European Theoretical Spectroscopy Facility (ETSF)}

\author{E. Khosravi}
\affiliation{Max-Planck Institut f\"ur Mikrostrukturphysik, Weinberg 2,
D-06120 Halle, Germany}
\affiliation{European Theoretical Spectroscopy Facility (ETSF)}

\author{C. Verdozzi}
\affiliation{Mathematical Physics, Lund University, 22100 Lund, Sweden}
\affiliation{European Theoretical Spectroscopy Facility (ETSF)}

\author{E.K.U. Gross}
\affiliation{Max-Planck Institut f\"ur Mikrostrukturphysik, Weinberg 2,
D-06120 Halle, Germany}
\affiliation{European Theoretical Spectroscopy Facility (ETSF)}

\date{\today}  

\begin{abstract}
The role of the discontinuity of the exchange-correlation potential of density functional theory is studied in the context of 
electron transport and shown to be intimately related to Coulomb blockade. 
By following the time evolution of 
an interacting nanojunction attached to biased leads, we find that, instead of evolving to a steady state, the system reaches 
a dynamical state characterized by correlation-induced current oscillations. Our results establish a dynamical picture of 
Coulomb blockade manifesting itself as a periodic sequence of charging and discharging of the nanostructure.
\end{abstract}
  
  
\maketitle  

Coulomb Blockade (CB) \cite{CB1} is one of the true hallmarks of 
electron-electron interactions in mesoscopic and nanoscale physics. 
In molecular transport, CB  is due to an electrostatic barrier 
induced by the electrons in the device which 
prevents further electrons from  tunneling in \cite{CB1b}. 
Progress in the theoretical description  and experimental manipulation of CB 
is expected to foster advances in nanoelectronics and 
quantum information technology \cite{CB2,maxbook,Neaton}. 
On the theoretical side, many aspects of CB can be understood in a rather 
simple way \cite{CB4}.
However, to achieve quantitative accuracy in real systems
an  {\it ab initio} description is required, 
something which has not been accomplished yet.

In fact,  most current {\it ab initio} treatments of transport are limited to 
the steady-state regime and based on the 
Landauer formalism combined with Density Functional Theory (DFT)
\cite{Lang:95}.
Within this prescription (L+DFT), the agreement with experiment is often poor, 
particularly for devices weakly coupled to leads. 
The approach has also been criticized on fundamental theoretical grounds 
(see e.g. \cite{KoentoppChangBurkeCar:08}). 
A main reason for its failure has been identified in the shortcomings of 
typical exchange-correlation (XC) functionals of static DFT 
\cite{SchmitteckertEvers:08}, such as, e.g., the lack of a derivative 
discontinuity in popular Local Density Approximations (LDA) and 
Generalized Gradient Approximations \cite{ToherFilippettiSanvitoBurke:05}. 
The importance of the derivative discontinuity in a static picture of CB for 
finite systems has been investigated in Ref.~[\onlinecite{CBKR}].

As transport is an intrinsically non-equilibrium phenomenon, the 
scientific community has progressively shifted to time-dependent (TD) 
approaches in recent years (see, e.g., Refs.~\cite{TDQT5,TDQT2}).
This shift also recognizes the fact that significant novel physics occurs in 
the time domain. While different approaches to TD transport  
have been developed,  TDDFT \cite{rg.1984}  offers 
a computationally efficient but still in principle exact 
method\cite{StefanucciAlmbladh:04}. 

In relation to CB, it is then quite natural to address 
the following 
issues:  i) what is the connection between a TDDFT 
description of CB in real time and the one of standard steady-state 
approaches? ii) which features an approximate XC  
functional should have to describe CB?

This Letter takes a first step in answering these basic questions.
We present here a TDDFT study of CB in the time domain for 
a correlated single-level quantum dot (QD) weakly coupled to leads. 
We use a novel XC functional \cite{LimaSilvaOliveiraCapelle:03} 
exhibiting the proper derivative discontinuity and
find that {\em the assumption that the system evolves towards a steady state 
is not generally justified}. In contrast, the discontinuity of the potential 
leads to self-sustained oscillations induced by electron correlations, with 
history-dependent frequencies and amplitudes. 
Our results thus reveal dynamical aspects of CB which are not accessible 
by traditional steady-state approaches.


{\it The system and its TDDFT time evolution. -} 
The Hamiltonian for a single-level QD coupled to two semi-infinite, 
non-interacting 1D leads is
\be\hat{H}(t)=
\hat{H}_{\rm 
QD}+\sum_{\alpha=L,R}\hat{H}_{\alpha}+\hat{H}_{T}+\hat{H}_{\rm bias}(t).
\label{hamil}
\ee
In Eq. (\ref{hamil}), $\hat{H}_{\alpha}=-\sum_{\sigma}\sum_{i=1}^{\infty} (V \hat{c}_{i+1 \alpha, 
\sigma}^{\dagger} \hat{c}_{i \alpha, \sigma} + {\rm h.c.})$, and
$\hat{H}_{T}=-\sum_{\alpha,\sigma} (V_{\rm link} \;
\hat{c}_{1 \alpha, \sigma}^{\dagger} \hat{c}_{0 \sigma} + {\rm 
h.c.})$ respectively describe, in standard notation, the leads and their 
coupling to the QD. $\hat{H}_{\alpha}$ and  $\hat{H}_T$
contain only nearest neighbor hopping terms: $V$  (in the left, $L$, and 
right, $R$, lead)  and $V_{\rm link}$ (to/from the QD). In the following 
we take both $V$ and $V_{\rm link}$ to be positive. The effect of applying a 
TD bias is contained in
$\hat{H}_{\rm bias}(t)=
\sum_{\alpha,\sigma}\sum_{i=1}^{\infty}W_{\alpha}(t) 
\hat{c}^{\dag}_{i \alpha, \sigma}\hat{c}_{i \alpha, \sigma}$.  
Finally, the QD Hamiltonian is 
\be
\hat{H}_{\rm QD}=
v_{\rm ext}  \sum_{\sigma}\hat{n}_{0\sigma} + U \hat{n}_{0\uparrow} 
\hat{n}_{0\downarrow}
\ee
with $\hat{n}_{0\sigma}=\hat{c}^{\dagger}_{0\sigma}\hat{c}_{0\sigma}$  the 
density for electrons with spin $\sigma$. 
Two parameters describe the QD: the charging energy $U$ and a static  
gate voltage $v_{\rm ext}$. For time $t\le 0$, the system is in equilibrium; 
at  $t > 0$, a bias $W_{\alpha}(t)$ is applied.

Within TDDFT, the many-body Hamiltonian (\ref{hamil}) is mapped onto 
an effective one-particle Kohn-Sham (KS) Hamiltonian.
The KS dynamics produces the exact TD density, provided we know the exact 
KS potential whose density dependence is non-local in space and time. 
In practice, one has to resort to approximations. 
As shown below, we can already gain significant insight via an approximate, 
adiabatic KS potential.  
The approximation we use is based on a ground state XC functional obtained from 
an LDA of the non-uniform 1D Hubbard model, via the Bethe Ansatz  
(BALDA)~\cite{LimaSilvaOliveiraCapelle:03}. In the context of TDDFT, an 
adiabatic version of this functional (ABALDA) has already been used to 
investigate the dynamics of Hubbard clusters \cite{Verdozzi:08}. 
Within the ABALDA, the KS Hamiltonian reads
$\hat{H}_{\rm KS}(t)=\hat{H}_{\rm QD, KS}(t)+
\sum_{\alpha=L,R}\hat{H}_{\alpha}+\hat{H}_{T}+\hat{H}_{\rm bias}(t)$,
with
\be
\hat{H}_{\rm QD, KS}(t)=\sum_{\sigma}v_{\rm KS}[n_0(t)] \hat{n}_{0\sigma}.
\ee
The KS potential is a functional of the instantaneous density 
$n_0(t)=\sum_{\sigma} 
n_{0\sigma}(t)$ on the QD. Explicitly, 
\be
v_{\rm KS}[n_0(t)] = v_{\rm ext} + \frac{1}{2} {U} n_0(t) + 
v_{\rm xc}[n_0(t)].\label{theKSpot}
\ee
The original BALDA 
functional was devised for a uniform chain with Hubbard 
interaction $U$ and nearest neighbor hopping $V$: in this case,  
the ratio $U/V$ is the relevant parameter in $v_{\rm xc}$ ~\cite{LimaSilvaOliveiraCapelle:03}. 
However, in the CB regime, which we wish to address here, the 
electrons need to be largely localized in the device.
For our model system, this corresponds to i) taking 
$V_{\rm link}$ significantly smaller than $V$ 
and ii) using ${U}/V_{\rm link}$ as relevant parameter in $v_{\rm xc}$. 
Accordingly, we propose the functional form
\bea
v_{\rm xc}[n]&=&\theta(1-n)v^{(<)}_{\rm xc}[n]-\theta(n-1)v^{(<)}_{\rm xc}[2-n],\hbox{\hspace{1.cm}}\label{vdisc}\\
v^{(<)}_{\rm xc}[n]&=&
-\frac{1}{2} {U} n - 2 V_{\rm link} \left[ 
\cos\left(\frac{\pi n}{2}\right) -
\cos\left(\frac{\pi n}{\beta}\right) \right],
\eea
where the parameter $\beta$ is determined by the condition
\be
\frac{2 \beta}{\pi} \sin(\pi/\beta) = 4 \int_0^{\infty} {\rm d} x \,
\frac{J_0(x) J_1(x)}{x [1 + \exp({U} x/(2 V_{\rm link}))]}
\label{eq_beta}
\ee
with $J_{i=0,1}(x)$ Bessel functions. For $V_{\rm link}\rightarrow 0$, 
$v_{\rm xc}[n] \rightarrow  -\theta(1-n){U} n /2 + \theta(n-1){U} (2-n) /2 $, 
which correctly reproduces the exact  XC potential of the isolated 
QD. This $v_{\rm xc}$ also leads to densities (not shown) in good 
agreement with the Quantum Monte Carlo data \cite{WangSpataruHybertsenMillis:08}
for the Anderson model. At half-filling, $v_{\rm xc}[n]$ is 
discontinuous \cite{LimaOliveiraCapelle:02}:  
$v_{\rm xc}[1^+] - 
v_{\rm xc}[1^-] =U - 4 V_{\rm link}  \cos(\frac{\pi}{\beta})$. 
The exact XC potential is certainly discontinuous for the 1-D 
Hubbard model or for the isolated QD. 
However, for the QD connected to non-interacting leads it is reasonable to expect a small
smoothening of the discontinuity (this is supported by preliminary
calculations, not shown here, of the exact $v_{\rm xc}$ for an Anderson 
impurity in small clusters).
We therefore modified  the 
discontinuous (at $n = 1$)  $v_{\rm xc}[n]$ of Eq. (\ref{vdisc}), with a softened, continuous 
version $ \tilde{v}_{\rm xc}(n) = f(n) v_{\rm xc}^{(<)}[n]
-(1-f(n)) v_{\rm xc}^{(<)}[2-n]$. Here, $f(n) = \frac{1}{\exp((n-1)/a) + 1}$,
with a smoothening parameter $a$. The smoothening of the KS potential 
also has the advantage of alleviating the numerical difficulties caused by 
the sudden changes of $v_{\rm xc}$  during time propagation.

\begin{figure}[t]
\includegraphics[width=0.47\textwidth]{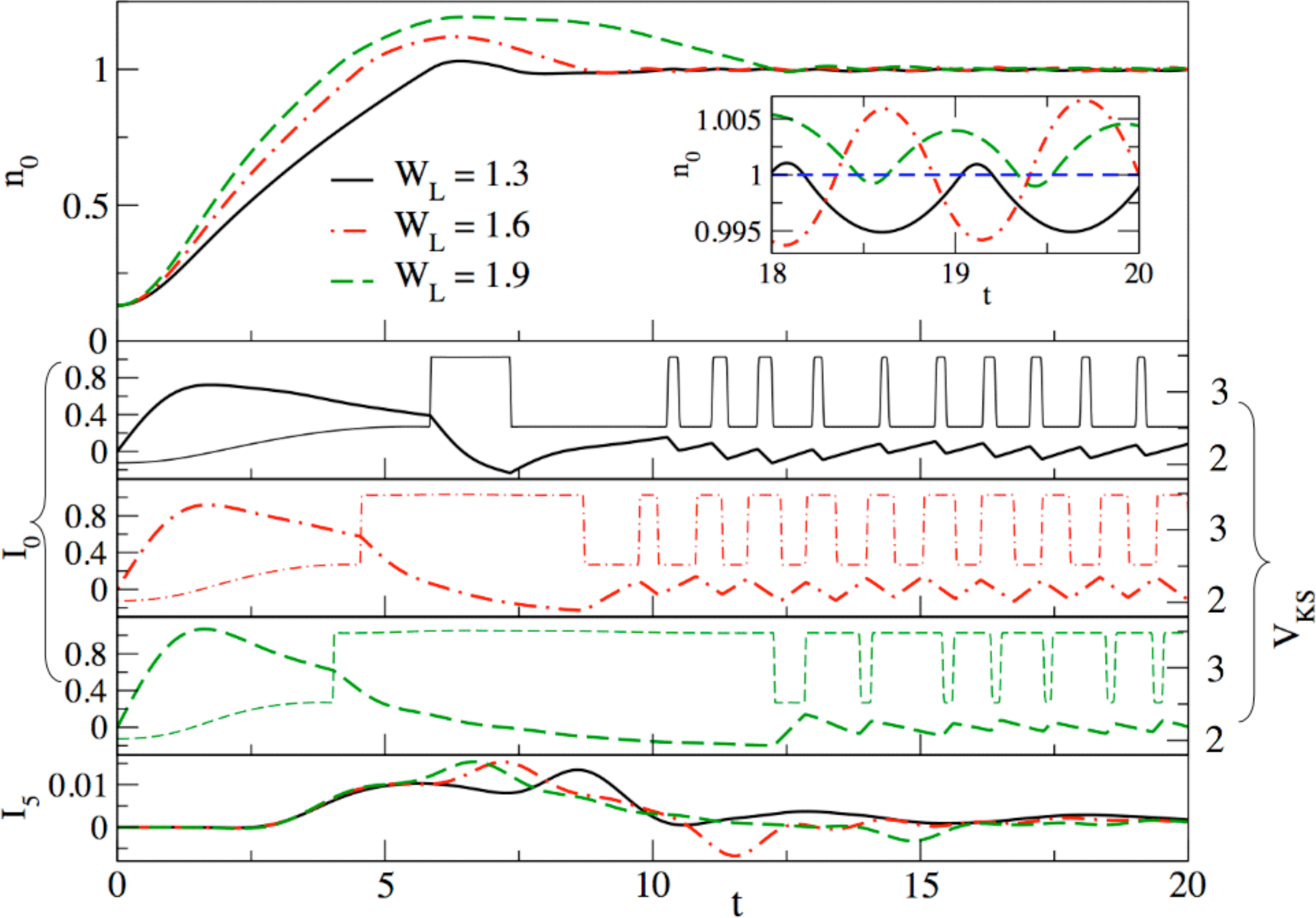}
\caption{ (Color online) Time evolution of the density, current and KS potential 
for three different biases. In all panels, solid, chain and dashed line refer
to $W_L=1.3,\, 1.6 ,\, 1.9$, respectively. Top panel: densities at the QD.
The inset shows the density at the end of the propagation period. 
Middle three panels: current through the QD (thick lines) and 
KS potential (thin lines). Bottom panel: current five sites away 
from the QD. }
\label{fanc}
\end{figure}

To propagate the lead-dot-lead KS system in time, we 
use a recently developed TD algorithm
\cite{KurthStefanucciAlmbladhRubioGross:05} for  
an open system of (effectively) noninteracting
electrons, as required in TDKS. 
In all simulations below, energies are measured in units of $V$, times 
in units of $V^{-1}$ and currents in units of $|e|V$ where $e$ is the 
charge of the carriers. 
The sharp slope near $n=1$ in $v_{\rm xc}$ has a profound  impact on the 
time evolution of the density on the QD as well as on the current through 
it. 

{\it TD transport in the CB regime.- } 
In Fig. \ref{fanc}, we show the time evolution of the density, the current  and 
the KS potential with a smoothening parameter $a=10^{-4}$ for three bias 
values and $V_{\rm link}=0.3$, $v_{\rm ext}=2$, ${U}=2$, 
and Fermi energy $\varepsilon_{\rm F}=  1.5$ (as shown below, this choice of 
the parameters corresponds to the CB regime).
The system is in its ground state for $t<0$; the equilibrium density
is calculated self-consistently with the $v_{\rm KS}$ of Eq. (\ref{theKSpot})
but in  static DFT. At $t=0$, a dc bias $W_L$ is suddenly switched on in the 
left lead. 
Remarkably, for this set of parameters and within a certain range of 
biases, {\em the system does not evolve towards a steady state};  instead, after
a transient period, we see self-sustained density oscillations
around unity (Fig. \ref{fanc}, top panel), 
with an amplitude of the order of  $10^{-3}$ (inset in the same panel).
For the lower  bias, $W_L = 1.3$, the density 
remains below unity for most of the period, for $W_L = 1.6$
the density roughly oscillates around 1, and for 
$W_L = 1.9$ the density exceeds unity for most of the period.
The oscillations are also present in the current (middle three panels), 
but this time with a much larger amplitude relative to 
its average value. The saw-tooth structure is consistent with the continuity 
equation (the density is approximately piece-wise parabolic).
Note that away from the QD,
the oscillations in the current (and in the density) disappear (bottom panel).
Oscillations are clearly visible also in $v_{\rm KS}$ (middle three panels). 
Due to the  fairly large jump in $v_{\rm xc}$ and the small amplitude of the 
density oscillations, the KS potential is a train of almost 
rectangular pulses. For larger biases the density remains above unity for 
longer times, and the width of the pulses in $v_{\rm KS}$ extends in time. 
The oscillations discussed here are a direct consequence of the 
discontinuity in $v_{\rm xc}$. For similar calculations (not shown) in the 
Hartree approximation, where $v_{\rm KS}$ is continuous in the density, the 
system does evolve towards a steady state.

{\it History dependence, oscillations and smoothening.-}
We have calculated the time-evolution of the system as obtained 
by switching the bias as $W_{L}(t)=W_{L}\sin^{2}(\frac{\pi t}{2T_{\rm switch}})$ 
for $t<T_{\rm switch}$ and $W_{L}(t) = W_L = {\rm const}$ otherwise.
The Fourier analysis of $n_0(t)$ 
reveals peaks whose position and height depend on $T_{\rm switch}$, 
i.e., on the history of the applied bias. We found that for any small 
but finite $a$, the amplitude of the oscillations 
approaches zero as $T_{\rm switch}\to \infty$. 
We therefore conclude that in this case the L+DFT approach gives the same 
solution as TDDFT for sufficiently slow switching \cite{TDremark}. 
For $a=0$, instead, the system never reaches a steady state.

{\it  CB regime from the steady-state approach.-} 
The oscillations just described are distinct from those occurring in the 
presence of single particle bound states \cite{s.2007,kksg.2009}: They are 
induced by electron correlations, i.e., are absent in the mean 
field (Hartree) approximation. An especially revealing feature is that at 
long times the system is in a {\em dynamical state of oscillating density 
and current}, whose time-average is fairly constant for a large range of 
biases. For further insight, we now use the L+DFT  approach which, as said 
above, yields the same solution as TDDFT for $a\neq 0$ and for adiabatic 
switching.
\begin{figure}[t]
\includegraphics[width=.47\textwidth]{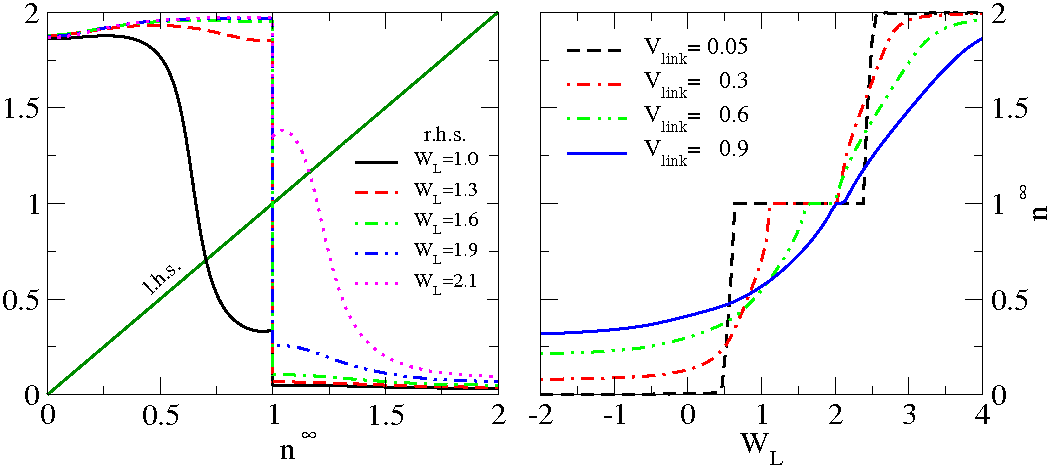}
\caption{(Color online) Left panel: Graphical solution of 
Eq. (\ref{dens_sc}) for few values of  the applied bias $W_{L}$. 
Right panel: Steady-state density as function of  $W_{L}$ for a smoothened 
$v_{\rm KS}$
($a=10^{-4}$) and few values of the dot-lead hopping $V_{\rm link}$.
}
\label{sc_dens}
\end{figure}
Using non-equilibrium Green's function  techniques,
the value $n^{\infty}$ of the steady-state density at the 
QD is given by the self-consistency condition\cite{addbound}
\be
n^{\infty } = 2 \sum_{\alpha=L,R} \int_{-\infty}^{\varepsilon_F+W_{\alpha}} 
\frac{{\rm d}\omega}{2 \pi} \Gamma(\omega-W_{\alpha}) |G(\omega)|^2  \; .
\label{dens_sc}
\ee
Here $G(\omega) = \left( \omega - v_{\rm KS}[n^{\infty}] - 
\sum_{\alpha} \Sigma(\omega-W_{\alpha})\right)^{-1}$
is the retarded Green function at the QD site, which depends on 
$n^{\infty}$ through $v_{\rm KS}$, $\Sigma(\omega)$ 
is the embedding self energy of the leads, and  
the width function $\Gamma(\omega)=-2{\rm Im}[\Sigma(\omega)]$. 

We first consider a discontinuous $v_{\rm xc}$ (i.e. $a=0$).
In Fig.~\ref{sc_dens}, left panel,  we display the l.h.s. and  the r.h.s. of 
Eq.~(\ref{dens_sc}) as function of $n^{\infty}$ for the 
parameters $V=1$, $V_{\rm link}=0.3$, $U=2$, $v_{\rm 
ext}=2$, $\varepsilon_F=1.5$ and for different values of 
the dc bias $W_{L}$. These are the parameters used in our TD
simulations of Fig.~\ref{fanc}. Interestingly, there is a range 
of applied bias voltages for which Eq.~(\ref{dens_sc}) 
{\em does not have a solution}. Since Eq.~(\ref{dens_sc}) is a 
condition for the QD density at the steady-state, 
it follows that, as   
a direct consequence of the discontinuity in $v_{\rm xc}$, for some values of $W_{\alpha}$, 
{\em no steady state exists}. On the other hand, for $a=10^{-4}$, 
as in our TD simulations, Eq.~(\ref{dens_sc}) 
admits steady state solutions; however, from Fig. \ref{fanc}, 
we know that, in general, the system time-evolves towards an oscillating 
regime and that the solution of the L+DFT scheme is a solution in TDDFT  
for adiabatic switching.

To show that in fact we {\it are} in the CB regime,  in Fig.~\ref{sc_dens}, right panel, 
we display the steady state density $n^{\infty}$, as a function of $W_L$. As 
before, $v_{\rm ext}=2$,  $\varepsilon_{\rm F}=1.5$, ${U}=2$ and $a=10^{-4}$. 
We see a plateau in the density at $n=1$ developing for a 
range of $W_L$.  The plateau increases with decreasing 
$V_{\rm link}$, and in the limit of small $V_{\rm link}$ 
its length becomes equal to the charging energy $U$.  This  is
the usual CB picture: if the site is occupied by one electron, 
a second electron can only jump in if its energy exceeds the 
charging energy of the QD.
For $V_{\rm link}=0.3$, the $W_L $'s of Fig.\ref{fanc} correspond to the beginning, middle, 
and end of the density plateau. Our results show that there exists a range 
of biases, starting at a critical value $W_c$, for which the decay time 
of the oscillations becomes infinite.

{\it Some considerations on $v_{xc}$ and the ABALDA.-}
Dropping only the assumption of locality in space, $v_{\rm xc}$ 
will depend on the instantaneous densities at few, say $M$, 
sites around the QD. 
This gives a steady state condition similar to Eq. (\ref{dens_sc}), 
in the form of a set of equations for the densities at the $M$ sites. 
For a discontinuous $v_{\rm xc}$ at one (or more) of the $M$ sites, 
these equations may not have a solution, similarly to
what we found in Fig. \ref{sc_dens}. 
The role of non-adiabaticity is much harder to assess since it 
requires the development of XC functionals with memory, e.g., 
via non-equilibrium Many-Body Perturbation Theory 
\cite{TDQT2,fva.2009,kristian:08}
or fluid dynamical considerations \cite{dbg.1997,t.2007}.

We can now provide an answer to the two questions posed at the beginning of 
this Letter. A discontinuous or rapidly varying $v_{\rm xc}$ is central to the 
description of CB within L+DFT and TDDFT.  
For KS potentials with a true discontinuity,  the steady-state 
self-consistency condition in L+DFT cannot always be satisfied. 
Considering a smoothened discontinuity (which, as said earlier, is physically 
more realistic) the L+DFT approach yields a clear-cut CB scenario.
However, the very same XC potential used in a TDDFT framework leads 
to a dynamical state with history-dependent, 
self-sustained density and current oscillations:  
where the static approach gives CB, the TD approach gives oscillations.
The average of these oscillations corresponds 
to the L+DFT solution. The latter is a  
TDDFT solution only for an adiabatic switch-on.

To conclude, our results suggest that in a transport setup CB manifests itself 
as an intrinsically TD phenomenon, i.e., it corresponds to an oscillatory 
current representing the intuitive picture of CB as a sequence of charging 
and discharging of the weakly coupled molecule or QD. While our calculations 
are performed for a model consisting of a single correlated QD 
coupled to non-interacting leads, we expect the results to be of general 
nature: In the continuum-real-space TDDFT description, whenever the particle 
number on the molecule or QD crosses an integer, the discontinuity of the
time-dependent XC potential \cite{LeinKuemmel:05} will trigger persisting
charge and current oscillations which cannot be captured in any steady-state 
approach.

We would like to acknowledge useful discussions with C\'{e}sar Proetto. 
SK, CV, and EKUG acknowledge the hospitality of KITP where part of this 
work was carried out. SK acknowledges funding by the 
``Grupos Consolidados UPV/EHU del Gobierno Vasco'' (IT-319-07).
CV is supported by ETSF (INFRA-2007-211956). This research was supported in 
part by the National Science Foundation under Grant No. PHY05-51164.


\begin{thebibliography}{10}

\bibitem{CB1} C.J. Gorter, Physica {\bf 17}, 777 (1951).

\bibitem{CB1b} I.O. Kulik and R.I. Shekhter, Zh. Eksp. Teor. Fiz. {\bf 68}, 623
(1975) [Sov. Phys. - JETP {\bf 41}, 308 (1975)].

\bibitem{CB2} See e.g. {\it Single Charge Tunneling: Coulomb Blockade 
Phenomena in Nanostructures}, H. Grabert and M. H. Devoret (eds.), NATO ASI 
Series B 234 (Plenum, New York, 1992). 

\bibitem{maxbook}
M. di Ventra, {\it Electrical Transport in Nanoscale Systems}, 
Cambridge University Press (2008).

\bibitem{Neaton}  See e.g. K. Phoa {\em et al.}, Nano Lett. 
{\bf 9}, 3225 (2009); S. Gustavsson {\em et al.}, Surf. Sci. Rep. {\bf 64}, 191 (2009).

\bibitem{CB4} C.W.J. Beenakker, Phys. Rev. B {\bf 44}, 1646 (1991).

\bibitem{Lang:95}
{N.D.~Lang}, Phys.~Rev.~B {\bf 52},  5335  (1995); 
J. Taylor {\em et al.}, Phys.~Rev.~B {\bf 63},  245407  (2001); 
M. Brandbyge {\em et al.},
Phys. Rev. B {\bf 65}, 165401 (2002); 
F. Evers {\em et al.}, Phys.~Rev.~B {\bf 69},  235411  (2004).

\bibitem{KoentoppChangBurkeCar:08}
M. Koentopp {\em et al.}, J.~Phys.~Condens.~Matter {\bf 20},
   083203  (2008).

\bibitem{SchmitteckertEvers:08}
P. Schmitteckert and F. Evers, Phys.~Rev. Lett. {\bf 100},  086401  (2008).

\bibitem{ToherFilippettiSanvitoBurke:05}
C. Toher {\em et al.}, Phys.~Rev. Lett. {\bf 95},
  146402  (2005).
  
\bibitem{CBKR} K. Capelle {\em et al.},   Phys.~Rev. Lett. {\bf 99}, 010402 (2007). 

\bibitem{TDQT5} 
R. Baer {\em et al.}, J. Chem. Phys. {\bf 120}, 3387 (2004); 
K. Burke {\em et al.}, Phys.~Rev. Lett. {\bf 94}, 146803 (2005); 
C. Verdozzi {\em et al.}, Phys. Rev. Lett. {\bf 97}, 046603 (2006); 
C.L. Cheng {\em et al.}, Phys.~Rev.~B {\bf 74},  155112  (2006); 
X. Zheng {\em et al.}, Phys. Rev. B {\bf 75}, 195127 (2007); 
P. Bokes {\em et al.}, Phys.~Rev. Lett. {\bf 101}, 046402 (2008); 
F. Heidrich-Meisner {\em et al.}, Phys. Rev. B {\bf 79}, 235336 (2009); 
V. Moldoveanu {\em et al.}, New J. Phys. {\bf 11}, 073019 (2009).

\bibitem{TDQT2} P. My\"oh\"anen {\em et al.}, Europhys. Lett. {\bf 84}, 67001 (2008);  Phys. Rev. B {\bf 80}, 115107 (2009).
   
\bibitem{rg.1984}
E. Runge and E.K.U.~Gross,
Phys. Rev. Lett. {\bf 52}, 997 (1984).

\bibitem{StefanucciAlmbladh:04}
G. Stefanucci and {C.-O.~Almbladh}, Europhys.~Lett. {\bf 67},  14  
(2004);  Phys.~Rev.~B {\bf 69},  195318  (2004).

\bibitem{LimaSilvaOliveiraCapelle:03}
{N.A.~Lima} {\em et al.}, Phys.~Rev. Lett. {\bf 90},  146402  (2003).

\bibitem{Verdozzi:08}
C. Verdozzi, Phys.~Rev. Lett. {\bf 101},  166401  (2008).

\bibitem{WangSpataruHybertsenMillis:08}
X. Wang {\em et al.}, Phys. Rev. B {\bf 77}, 045119 (2008).

\bibitem{LimaOliveiraCapelle:02}
{N.A.~Lima} {\em et al.}, Europhys.~Lett. {\bf 60},  601
  (2002).

\bibitem{KurthStefanucciAlmbladhRubioGross:05}
S. Kurth {\em et al.}, Phys.~Rev.~B {\bf 72},  035308  (2005).
  
\bibitem{TDremark} 
The steady-state solution achieved in TDDFT with an 
adiabatically switched bias 
is the same as the L+DFT one only for adiabatic 
approximations to $v_{\rm xc}$
\cite{StefanucciAlmbladh:04,KBE:06,transrev}. 

\bibitem{KBE:06}
M. Koentopp {\em et al.}, Phys. Rev. B {\bf 73}, 121403(R) (2006). 

\bibitem{transrev} 
G. Stefanucci {\em et al.}, in: {\it Molecular and 
Nano Electronics: Analysis, Design, and Simulation}, ed. J.~Seminario 
(Elsevier, Amsterdam, 2006)

\bibitem{s.2007}
G. Stefanucci,
Phys. Rev. B {\bf 75}, 195115 (2007). 

\bibitem{kksg.2009}
E. Khosravi {\em et al.}, Appl.~Phys.~A {\bf 93}, 355 (2008);  
Phys. Chem. Chem. Phys. {\bf 11}, 4535 (2009). 

\bibitem{addbound} Another term must be added on the r.h.s. of Eq. (\ref{dens_sc}) 
if the asymptotic $t\rightarrow \infty$ Hamiltonian has split-off 
eigenstates \cite{s.2007,kksg.2009}. 

  
\bibitem{fva.2009}
M. Puig von Friesen, C. Verdozzi, and C.-O. Almbladh, 
Phys. Rev. Lett. {\bf 103}, 176404 (2009). 
\bibitem{kristian:08}
K.S. Thygesen, Phys. Rev. Lett. {\bf 100}, 166804 (2008). 

\bibitem{dbg.1997}
J.F. Dobson {\em et al.}, 
Phys. Rev. Lett. {\bf 79}, 1905 (1997).

\bibitem{t.2007}
I.V. Tokatly, 
Phys. Rev. B {\bf 75}, 125105 (2007), and references therein.

\bibitem{LeinKuemmel:05}
M. Lein, S. K\"ummel, Phys. Rev. Lett. {\bf 94}, 143003 (2005). 

\end{thebibliography}
\end{document}